\documentclass[aps,prl,preprint,groupedaddress]{revtex4}

\begin{document}

\preprint{HU-EP-04/25}

\title{
Confinement in the Abelian-Higgs--type theories:
string picture and field correlators} \thanks{Invited contribution to the collection 
of articles devoted to the 70th birthday of Yu.A. Simonov.}

\author{Dmitri Antonov}
\email{antonov@physik.hu-berlin.de}
\thanks{Permanent address:
ITEP, 
B. Cheremushkinskaya 25, RU-117 218 Moscow, Russia.} 

\author{Dietmar Ebert} 
\email{debert@physik.hu-berlin.de}

\affiliation{Institute of Physics, Humboldt University of Berlin,
Newtonstr. 15, 12489 Berlin, Germany}

\date{\today}

\begin{abstract}
Field correlators and the string representation are used as two complementary approaches 
for the description of confinement in the SU($N$)-inspired dual Abelian-Higgs--type model.
In the London limit of the simplest, SU(2)-inspired, model,
bilocal electric field-strength correlators have been derived with accounting for the contributions to these  
averages produced by closed dual strings. The Debye screening in the plasma of such strings 
yields a novel long-range interaction between points lying on the contour of the Wilson loop.
This interaction generates a L\"uscher-type term, even when one restrics
oneself to the minimal surface, as it is usually done in the bilocal approximation to
the stochastic vacuum model. Beyond the London limit, it has been shown that
a modified interaction appears, which becomes reduced to the standard Yukawa one in the 
London limit. Finally, a string representation of the SU($N$)-inspired model with the 
$\Theta$-term, in the London limit, can be constructed. 
\end{abstract}

\pacs{}

\maketitle

\section{Introduction}

The Stochastic Vacuum Model (SVM)~\cite{1} is nowadays commonly 
recognized as a promising nonperturbative approach to QCD
(see Ref.~\cite{2} for reviews).
Within the so-called bilocal or Gaussian approximation, 
well confirmed by the existing lattice data~\cite{3,4}, this model 
is fully described by the irreducible
bilocal gauge-invariant field strength correlator (cumulant),
$\left<\left<F_{\mu\nu}(x)\Phi(x,x')
F_{\lambda\rho}(x')\Phi(x',x)\right>\right>$. Here, 
$F_{\mu\nu}=\partial_\mu A_\nu-\partial_\nu A_\mu-ig[A_\mu, A_\nu]$
stands for the Yang-Mills field-strength tensor, $\Phi(x,y)\equiv
\frac{1}{N_c}{\cal P}\exp\left(ig\int\limits_{y}^{x}
A_\mu(u)du_\mu\right)$ is a 
parallel transporter factor along the straight-line
path, and $\left<\left<{\cal O}{\cal O}'\right>\right>\equiv
\left<{\cal O}{\cal O}'\right>-\left<{\cal O}\right>\left<{\cal O}'\right>$
with the average defined {\it w.r.t.} the Euclidean Yang-Mills action.
It is further convenient to parametrize the bilocal cumulant by 
the two coefficient functions $D$ and $D_1$~\cite{1,2} as follows:

$$
\frac{g^2}{2}\left<\left<F_{\mu\nu}(x)\Phi(x,x')
F_{\lambda\rho}(x')\Phi(x',x)\right>\right>=\hat 1_{N_c\times 
N_c}\left\{(\delta_{\mu\lambda}\delta_{\nu\rho}-\delta_{\mu\rho}
\delta_{\nu\lambda})D\left((x-x')^2\right)+\right.$$

\begin{equation}
\label{dd1}
\left.+\frac12\left[\partial_\mu^x((x-x')_\lambda
\delta_{\nu\rho}-(x-x')_\rho\delta_{\nu\lambda})+
\partial_\nu^x((x-x')_\rho\delta_{\mu\lambda}-(x-x')_\lambda
\delta_{\mu\rho})\right]D_1\left((x-x')^2\right)\right\}.
\end{equation}
After that, setting for the nonperturbative parts of the $D$- and 
$D_1$-function various {\it Ans\"atze}, one can apply SVM to 
calculations of the high-energy scattering processes~\cite{5}
or test these {\it Ans\"atze} in the lattice experiments~\cite{3,4}.
However, from the pure field-theoretical point of view, 
a challenge remains to derive 
the coefficient functions analytically. Unfortunately, in QCD, this problem looks too
complicated.

To proceed with, it is therefore reasonable to derive field-strength correlators not in QCD itself,
but rather in some Abelian-type QCD-inspired models, which inherit confinement and allow for its 
analytic description. 
These include SU(2)-~\cite{th} and 
SU(3)- \cite{suz} inspired dual Abelian-Higgs--type theories, as well as 3D compact QED~\cite{comp}.
The bilocal field-strength cumulant in these theories has been studied 
in Refs.~\cite{7,8,9}, respectively.
In the present minireview, we will briefly survey the results which concern the dual Abelian-Higgs--type theories, 
as well as their further 
elaborations performed in Ref.~\cite{10}. For the sake of simplicity, we will restrict ourselves to the 
SU(2)-inspired case, {\it i.e.}, a simple dual Abelian-Higgs model (DAHM), 
although the SU(3)-generalization is straightforward~\cite{8}.

One important fact for the further discussion is that in DAHM a sector with closed dual strings~\cite{11} exists.
Such closed strings are short-living (virtual) objects, whose typical sizes are much smaller than the 
typical distances between them. This means that, similarly to 
monopoles in 3D compact QED, closed strings can be treated in the dilute-plasma approximation.
Moreover, in the leading (semi-classical) approximation, the interaction of closed dual strings 
with large open ones, which end up at external quarks, can be disregarded at all. This is 
precisely the approximation in which field-strength correlators have been evaluated in Refs.~\cite{7,8}.
A leading correction to these semi-classical expressions, which 
stems from the interaction of closed strings with the open ones, has been found in Ref.~\cite{10} and 
will be reviewed below.

The outline of the minireview is as follows. In the next Section, we will first mention a correspondence,
based on the Abelian-projection method, 
between the DAHM and the SU(2)-QCD, which will be needed for the 
future purposes. Secondly, we will briefly review the main 
results of a calculation of electric field-strength correlators in the approximation when 
closed strings are disregarded. In the subsequent Section, 
after a brief review of properties of the grand canonical ensemble 
of closed strings, we will consider the contribution of these objects to the 
field-strength correlators. In the same Section, we will also discuss two types of corrections to the $\bar qq$-potential -
due to closed strings and due to the deviation from the London limit. In the last Section, we will present a string representation of the 
SU($N$)-inspired analogue of DAHM extended by the $\Theta$-term.
The main results will finally be quoted in Summary.

\section{Electric field-strength correlators in the absence 
of closed strings}

\subsection{The model}

To derive from the  
Lagrangian of the SU(2)-gluodynamics an IR effective
theory, based on the assumption of condensation of 
Abelian-projected monopoles, one usually employs the so-called 
Abelian dominance hypothesis~\cite{15}. It states that the off-diagonal 
(in the sense of the Cartan decomposition) fields can be disregarded, since 
after the Abelian projection those can be shown to become very heavy
and therefore irrelevant to the IR region.
The action describing the remaining diagonal fields and Abelian-projected
monopoles reads

\begin{equation}
\label{et}
S_{\rm eff.}\left[a_\mu, f_{\mu\nu}^{\rm m}\right]=
\frac14\int d^4x\left(f_{\mu\nu}+
f_{\mu\nu}^{\rm m}
\right)^2. 
\end{equation}
Here, $a_\mu\equiv A_\mu^3$, $f_{\mu\nu}=\partial_\mu a_\nu-\partial_\nu
a_\mu$, and 
the monopole field-strength tensor $f_{\mu\nu}^{\rm m}$
obeys Bianchi identities modified by monopoles, 
$\partial_\mu\tilde f^{\rm m}_{\mu\nu}
\equiv
\frac12\varepsilon_{\mu\nu\lambda\rho}
\partial_\mu f^{\rm m}_{\lambda\rho}=j_\nu^{\rm m}$.
The monopole currents $j_\mu^{\rm m}$'s should eventually 
be averaged over in the sense, which will be specified below.

To proceed with the investigation of the monopole ensemble, 
it is useful to dualize 
the theory under study. This yields the following expression
for the partition function:

\begin{equation}
\label{et2}
{\cal Z}=\left<\int {\cal D}B_\mu
\exp\left[-\int d^4x\left(\frac14F_{\mu\nu}^2
-iB_\mu j_\mu^{\rm m}\right)\right]\right>_{j_\mu^{\rm m}},
\end{equation}
where $B_\mu$ is the magnetic vector-potential dual to the 
electric one, $a_\mu$, and $F_{\mu\nu}=\partial_\mu B_\nu-
\partial_\nu B_\mu$. Once the $j_\mu^{\rm m}$-dependence of the 
action became explicit, it is now possible to set up the 
properties of the monopole ensemble. To describe the condensation 
of monopoles, it is first necessary to specify $j_\mu^{\rm m}$
as the collective current of $N$ of those:
$j_\mu^{{\rm m}{\,}(N)}(x)=g_m\sum\limits_{n=1}^{N}\oint dx_\mu^n(s)
\delta(x-x^n(s))$.
Here, the world line of the $n$-th monopole is parametrized 
by the vector $x_\mu^n(s)$, and 
$g_m$ is the magnetic coupling constant, related to the QCD
coupling constant $g$ via the quantization condition 
$gg_m=4\pi n$ with $n$ being an integer. In what follows, 
we will for concreteness restrict ourselves to the monopoles possessing
the minimal charge, {\it i.e.} set $n=1$, although the generalization 
to an arbitrary $n$ is straightforward.
Further, it is necessary to set for the measure 
$\left<\ldots\right>_{j_\mu^{\rm m}}$ the following expression~\cite{16}:

$$
\left<\exp\left(i\int d^4xB_\mu j_\mu^{\rm m}\right)\right>_{j_\mu^{\rm m}}=
1+\sum\limits_{N=1}^{\infty}\frac{1}{N!}\left[\prod\limits_{n=1}^{N}
\int\limits_{0}^{\infty}\frac{ds_n}{s_n}{\rm e}^{2\lambda\eta^2s_n}
\int\limits_{u(0)=u(s_n)}^{}{\cal D}u(s_n')\right]\times$$

\begin{equation}
\label{et3}
\times\exp\left\{\sum\limits_{l=1}^{N}\int\limits_{0}^{s_l}ds_l'\left[
-\frac14\dot u^2(s_l')+ig_m\dot u_\mu(s_l')B_\mu(u(s_l'))\right]-
\lambda\sum\limits_{l,k=1}^{N}\int\limits_{0}^{s_l}ds_l'
\int\limits_{0}^{s_k}ds_k''\delta\left[u(s_l')-u(s_k'')\right]\right\}.
\end{equation}
Here, the vector $u_\mu(s_n')$ parametrizes the same contour as the 
vector $x_\mu^n(s)$. Clearly, the world-line action standing in the 
exponent on the R.H.S. of Eq.~(\ref{et3}) contains besides the usual 
free part also the term responsible for the short-range repulsion (else called self-avoidance) of the 
trajectories of monopoles.
Equation~(\ref{et3}) can further be rewritten as an integral over the 
dual Higgs field as follows:

\begin{equation}
\label{et4} 
\left<\exp\left(i\int d^4xB_\mu j_\mu^{\rm m}\right)
\right>_{j_\mu^{\rm m}}=
\int {\cal D}\Phi {\cal D}\Phi^{*}
\exp\left\{-\int d^4x\left[\left|D_\mu
\Phi\right|^2+\lambda\left(|\Phi|^2-\eta^2\right)^2\right]\right\},
\end{equation}
where $D_\mu=\partial_\mu-ig_mB_\mu$ is the covariant 
derivative. Finally, substituting 
Eq.~(\ref{et4}) into Eq.~(\ref{et2}), we arrive at the DAHM:

\begin{equation}
\label{et5}
{\cal Z}=\int \left|\Phi\right| {\cal D}\left|\Phi\right|
{\cal D}\theta {\cal D}B_\mu\exp\left\{-\int d^4x\left[
\frac14 F_{\mu\nu}+\left|D_\mu
\Phi\right|^2+\lambda\left(|\Phi|^2-\eta^2\right)^2\right]\right\},
\end{equation}
where $\Phi(x)=\left|\Phi(x)\right|{\rm e}^{i\theta(x)}$. 
The masses of the dual vector boson and of the dual Higgs field, derivable upon the 
substitution $\Phi(x)=\eta+\frac{\varphi(x)}{\sqrt{2}}$, read $m_B\equiv m=\sqrt{2}g_m\eta$ and $m_H=2\eta\sqrt{\lambda}$, 
respectively.
Clearly, the two main assumptions, 
made in course of this derivation, were 
the neglection of the off-diagonal degrees of freedom and  
the postulate that the monopole condensate can be modeled by the dual Higgs field.

\subsection{Bilocal electric field-strength correlator}

In order to investigate the bilocal cumulant of electric field 
strengths in the model~(\ref{et5}), it is necessary
to extend this model by external electrically charged test particles
[{\it i.e.} particles, charged {\it w.r.t.} the Cartan subgroup
of the original SU(2)-group]. It is therefore natural to call 
these particles simply ``quarks''. 
Such an extension can be performed by adding to the 
action~(\ref{et}) the term $i\int d^4xa_\mu j_\mu^{\rm e}$ with  
$j_\mu^{\rm e}(x)\equiv g\oint
\limits_{C}^{}dx_\mu(s)\delta(x-x(s))$ standing for the conserved 
electric current of a quark, which moves along a certain closed contour $C$.
Then, performing the dualization of the so-extended action
and summing up over 
monopole currents according to Eq.~(\ref{et3}), 
we arrive at Eq.~(\ref{et5}) with $F_{\mu\nu}$ replaced by
$F_{\mu\nu}+F_{\mu\nu}^{\rm e}$. Here,  
$F_{\mu\nu}^{\rm e}$ stands for the field-strength tensor 
generated by quarks according to the equation $\partial_\mu\tilde 
F_{\mu\nu}^{\rm e}=j_\nu^{\rm e}$. 
A solution to this equation reads $F_{\mu\nu}^{\rm e}
=-g\tilde\Sigma_{\mu\nu}^{\rm e}$, where   
$\Sigma_{\mu\nu}^{\rm e}(x)\equiv\int\limits_{\Sigma^{\rm e}}^{}
d\sigma_{\mu\nu}(\bar x(\xi))\delta(x-\bar x(\xi))$ 
is the so-called vorticity 
tensor current defined at an arbitrary surface $\Sigma^{\rm e}$
(which is just the world sheet of an open dual 
Nielsen-Olesen string), bounded by the contour $C$, and 
$\xi$ is a 2D-coordinate.

From now on, we will be interested 
in the London limit of DAHM, $\lambda\to\infty$, where it admits an exact 
string representation. In that limit, the partition function~(\ref{et5}) with external quarks reads

\begin{equation}
\label{vosem}
{\cal Z}=\int {\cal D}B_\mu {\cal D}\theta
\exp\left\{-\int d^4x\left[\frac14
\left(F_{\mu\nu}+F_{\mu\nu}^{\rm e}\right)^2+\eta^2
\left(\partial_\mu\theta-g_mB_\mu\right)^2\right]\right\}. 
\end{equation}
In Eq.~(\ref{vosem}), one next performs a 
decomposition of the phase of the dual 
Higgs field $\theta=
\theta^{{\rm sing.}}+\theta^{{\rm reg.}}$, where the multivalued field 
$\theta^{{\rm sing.}}(x)$ 
describes a certain configuration of dual strings and 
obeys the equation~\cite{17, more17}
 
\begin{equation}
\label{devyat}
\varepsilon_{\mu\nu\lambda\rho}\partial_\lambda
\partial_\rho\theta^{{\rm sing.}}(x)=2\pi\Sigma_{\mu\nu}(x), 
\end{equation}
and the integration measure becomes factorized, ${\cal D}\theta={\cal D}\theta^{\rm sing.}{\cal D}\theta^{\rm reg.}$.
Here, $\Sigma_{\mu\nu}$ stands for the 
vorticity tensor current, defined at the world sheet $\Sigma$ of a 
closed dual string, parametrized by the vector $x_\mu(\xi)$.
On the other hand, the field $\theta^{\rm reg.}(x)$ 
describes simply a singlevalued fluctuation around the above-mentioned
string configuration. Note that Eq.~(\ref{devyat}) is nothing, but the 
Stokes' theorem for $\partial_\mu\theta^{\rm sing.}$, written in the local form.

The string representation of the theory~(\ref{vosem}) 
can be derived analogously to Ref.~\cite{17}, where this has been done 
for a model with a global U(1)-symmetry. One obtains

\begin{equation}
\label{odinnad}
{\cal Z}=
\int {\cal D}x_\mu(\xi) {\cal D}h_{\mu\nu} 
\exp\Biggl\{-\int d^4x\left[\frac1{24\eta^2}H_{\mu\nu
\lambda}^2+\frac{g_m^2}{4}h_{\mu\nu}^2+i\pi h_{\mu\nu}\hat\Sigma_{\mu\nu}
\right]\Biggr\},
\end{equation}
where $\hat\Sigma_{\mu\nu}
\equiv 2\Sigma_{\mu\nu}^{\rm e}-\Sigma_{\mu\nu}$, and   
$H_{\mu\nu\lambda}\equiv\partial_\mu h_{\nu\lambda}+
\partial_\lambda h_{\mu\nu}+\partial_\nu h_{\lambda\mu}$ is the field-strength 
tensor of a massive antisymmetric spin-1 tensor field $h_{\mu\nu}$. This field 
emerged as a solution of some constraints arising from the integration over 
$\theta^{\rm reg.}$ and represents the  
massive dual vector boson. As far as the integration over the 
world sheets of closed strings, 
$\int {\cal D}x_\mu(\xi)$, is concerned, it appeared from the 
integration over $\theta^{\rm sing.}$ by virtue of  
Eq.~(\ref{devyat}), which established a one-to-one 
correspondence between $\theta^{\rm sing.}$ and $x_\mu(\xi)$. 
Physically this correspondence stems from the fact that the singularity 
of the phase of the dual Higgs field takes place just at closed-string 
world sheets. [Notice that, since in what follows we will be 
interested in effective actions, rather than the integration measures, 
the Jacobian emerging during the change of the integration variables 
$\theta^{\rm sing.}\to x_\mu(\xi)$, which has been evaluated 
in Ref.~\cite{19}, will not be discussed below and is assumed 
to be included in the measure ${\cal D}x_\mu(\xi)$.]

Finally, the Gaussian 
integration over the field $h_{\mu\nu}$ in Eq.~(\ref{odinnad}) 
leads to the following expression for the  
partition function~(\ref{vosem}):

$$
{\cal Z}=\exp\left[-\frac{g^2}{2}\oint\limits_C^{}dx_\mu
\oint\limits_C^{}dy_\mu D_m^{(4)}(x-y)\right]\times$$

\begin{equation}
\label{pyatnad}
\times\int {\cal D}x_\mu(\xi)\exp\left[-2(\pi\eta)^2\int d^4x\int d^4y
\hat\Sigma_{\mu\nu}(x)D_m^{(4)}(x-y)\hat\Sigma_{\mu\nu}(y)
\right]. 
\end{equation}
Here, $D_m^{(4)}(x)\equiv mK_1(m|x|)/(4\pi^2|x|)$ is the 
propagator of the dual vector boson, and $K_\nu$'s
henceforth stand for the modified Bessel functions. 
Clearly, the first exponential factor 
on the R.H.S. of Eq.~(\ref{pyatnad}) 
is the standard result, which can be obtained without accounting for the 
dual Nielsen-Olesen strings. Contrary to that, 
the integral over string world sheets on the R.H.S. of 
that equation stems just from the contribution of closed strings to the 
partition function and is 
the essence of the string representation.
The respective string effective action describes
both the interaction of closed world sheets $\Sigma$'s with the  
open world sheets $\Sigma^{\rm e}$'s and self-interactions of these
objects.

We are now in the position to discuss the bilocal correlator 
of electric field strengths in the model~(\ref{vosem}).
Indeed, owing to the Stokes' theorem,
such an extended partition function (which is actually nothing,
but the Wilson loop of a test quark) 
can be written as $\left<\exp\left(-\frac{ig}{2}\int d^4x
\Sigma^{\rm e}_{\mu\nu}f_{\mu\nu}\right)
\right>_{a_\mu, j_\mu^{\rm m}}$, where 
$\left<\ldots\right>_{a_\mu, j_\mu^{\rm m}}\equiv\left<\int 
{\cal D}a_\mu\exp
\left(-S_{\rm eff.}\left[a_\mu, f_{\mu\nu}^{\rm m}\right]\right) 
\left(\ldots\right)\right>_{j_\mu^{\rm m}}$ with 
$S_{\rm eff.}$ and $\left<\ldots\right>_{j_\mu^{\rm m}}$
given by Eqs.~(\ref{et}) and (\ref{et3}), respectively. Applying to this 
expression the cumulant expansion, we have in the bilocal approximation:

\begin{equation}
\label{Zonehand}
{\cal Z}\simeq\exp\left[-\frac{g^2}{8}\int d^4x\int d^4y
\Sigma_{\mu\nu}^{\rm e}(x)\Sigma_{\lambda\rho}^{\rm e}(y)\left<\left<
f_{\mu\nu}(x)f_{\lambda\rho}(y)
\right>\right>_{a_\mu, j_\mu^{\rm m}}\right].
\end{equation}
Following the SVM, let us parametrize the bilocal cumulant 
$\left<\left<f_{\mu\nu}(x)f_{\lambda\rho}(0)\right>\right>$
similarly to the parametrization of Eq.~(\ref{dd1}), namely set for this 
quantity the following {\it Ansatz}:

\begin{equation}
\label{dvaddva}
\Biggl(\delta_{\mu\lambda}\delta_{\nu\rho}-\delta_{\mu\rho}
\delta_{\nu\lambda}\Biggr){\cal D}\left(x^2\right)+
\frac12\Biggl[\partial_\mu
\Biggl(x_\lambda\delta_{\nu\rho}-x_\rho\delta_{\nu\lambda}\Biggr)
+\partial_\nu\Biggl(x_\rho\delta_{\mu\lambda}-x_\lambda\delta_{\mu\rho}
\Biggr)\Biggr]{\cal D}_1\left(x^2\right). 
\end{equation}
Owing to the Stokes' theorem, Eq.~(\ref{dvaddva}) yields

\begin{equation}
\label{eventual}
{\cal Z}\simeq\exp\left\{-\frac18\int d^4x\int d^4y\left[2g^2
\Sigma_{\mu\nu}^{\rm e}(x)
\Sigma_{\mu\nu}^{\rm e}(y)
{\cal D}\left((x-y)^2\right)+
j_\mu^{\rm e}(x)j_\mu^{\rm e}(y)\int\limits_{(x-y)^2}^{\infty}dt
{\cal D}_1(t)\right]\right\}.
\end{equation} 

On the other hand, Eq.~(\ref{eventual}) should coincide with 
Eq.~(\ref{pyatnad}) divided by ${\cal Z}\left[\Sigma_{\mu\nu}^{\rm e}=0
\right]$ (that is just the standard normalization 
condition, encoded in the integration measures), {\it i.e.} it reads

$${\cal Z}=\exp\left\{-\int d^4x\int d^4y D_m^{(4)}(x-y)\left[
8(\pi\eta)^2\Sigma_{\mu\nu}^{\rm e}(x)
\Sigma_{\mu\nu}^{\rm e}(y)+\frac12
j_\mu^{\rm e}(x)j_\mu^{\rm e}(y)\right]\right\}\times$$

\begin{equation}
\label{otherhand}
\times\left<\exp\left[8(\pi\eta)^2\int d^4x\int d^4y
D_m^{(4)}(x-y)\Sigma_{\mu\nu}^{\rm e}(x)
\Sigma_{\mu\nu}(y)\right]\right>_{x_\mu(\xi)},
\end{equation}
where the average $\left<\ldots\right>_{x_\mu(\xi)}$ is defined {\it w.r.t.} the action 
$2(\pi\eta)^2\int d^4x\int d^4y\Sigma_{\mu\nu}(x)D_m^{(4)}(x-y)\Sigma_{\mu\nu}(y)$.
As it has already been discussed in the Introduction, in the semi-classical approximation,
closed dual strings can be disregarded, since their typical areas $|\Sigma|$'s
are much smaller than the area $|\Sigma^{\rm e}|$ 
of the world sheet of a long open string, which confines a test quark.
Owing to this fact, the exponential factor, which should be averaged over 
closed strings on the R.H.S. of Eq.~(\ref{otherhand}), may be 
disregarded {\it w.r.t.} the first exponential factor in that equation, 
as well. Then, the comparison 
of the latter one with Eq.~(\ref{eventual}) readily yields 
for the functions ${\cal D}$ and ${\cal D}_1$ the following expression

\begin{equation}
\label{dvadtri}
{\cal D}\left(x^2\right)=\frac{m^3}{4\pi^2}
\frac{K_1(m|x|)}{\left|x\right|},   
\end{equation}

\begin{equation}
\label{dvadchetyr}
{\cal D}_1\left(x^2\right)=
\frac{m}{2\pi^2x^2}\Biggl[\frac{K_1(m|x|)}{\left|x\right|}
+\frac{m}{2}\Biggl(K_0(m|x|)+K_2(m|x|)\Biggr)\Biggr]. 
\end{equation}
In the IR limit, $\left|x\right|\gtrsim m^{-1}$, 
the asymptotic behaviours of the coefficient functions~(\ref{dvadtri}) 
and~(\ref{dvadchetyr}) 
are given by  

\begin{equation}
\label{dvadpyat}
{\cal D}\longrightarrow\frac{m^4}{4\sqrt{2}\pi^{\frac32}}
\frac{{\rm e}^{-m\left|x\right|}}{\left(m\left|x\right|\right)^
{\frac32}}  
\end{equation}
and 

\begin{equation}
\label{dvadshest}
{\cal D}_1\longrightarrow\frac{m^4}{2\sqrt{2}\pi^{\frac32}}
\frac{{\rm e}^{-m\left|x\right|}}{\left(m\left|x\right|\right)^
{\frac52}}. 
\end{equation}

One can now see that, according to the lattice data~\cite{3,4},  
the asymptotic behaviours~(\ref{dvadpyat}) and~(\ref{dvadshest}) 
are very similar 
to the IR ones of the nonperturbative parts of the 
functions $D$ and $D_1$, which parametrize the bilocal 
cumulant~(\ref{dd1}) in QCD.
In particular, both 
functions decrease exponentially, and the function ${\cal D}$ 
is much larger 
than the function ${\cal D}_1$ 
due to the preexponential power-like behaviour.
We also see that the role of the correlation length of the 
vacuum, $T_g$, 
{\it i.e.} the distance at which the functions $D$ and $D_1$ decrease,
is played in the model~(\ref{vosem}) by the inverse 
mass of the dual vector boson, $m^{-1}$.

Hence we see that, within the approximation when the contribution of 
closed strings to the partition function~(\ref{otherhand}) is disregarded, 
the bilocal approximation to the SVM is an exact
result in the theory~(\ref{vosem}),
{\it i.e.} all the cumulants
of the orders higher than the second one vanish. Higher cumulants
naturally appear upon performing in Eq.~(\ref{otherhand})
the average over closed strings. However, this average 
yields important modifications already on the level of the bilocal cumulant. 
Namely, as we will see in the next Section, 
it modifies the semi-classical expressions~(\ref{dvadtri}) 
and~(\ref{dvadchetyr}).

\section{Corrections to the $\bar QQ$-potential produced by closed strings and a finite Higgs mass}

To study the properties of closed strings, it is enough to consider the theory without 
external quarks.
The field-strength correlators can be addressed afterwards, {\it i.e.} 
already after
the summation over the grand canonical ensemble of closed strings.
Thus, let us first consider the theory~(\ref{vosem}) with 
$F_{\mu\nu}^{\rm e}=0$. Upon the derivation of the  
string representation of such a theory, we are then left
with Eq.~(\ref{odinnad}), where $\Sigma_{\mu\nu}^{\rm e}=0$.
To study the grand canonical ensemble of closed strings, it is 
necessary to replace $\Sigma_{\mu\nu}$ in Eq.~(\ref{odinnad}) 
by the following expression: 
$\Sigma_{\mu\nu}^N(x)=\sum\limits_{i=1}^{N}n_i\int 
d\sigma_{\mu\nu}(x_i(\xi))\delta(x-x_i(\xi))$. Here, 
$n_i$'s stand for winding numbers. In what follows, we will 
restrict ourselves to closed strings possessing the minimal
winding numbers, $n_i=\pm 1$. That is because, analogously to the 
3D-case~\cite{11,22}, the energy 
of a single closed string is known to be a quadratic function 
of its flux, owing to which the vacuum prefers to maintain two closed strings
of a unit flux, rather than one string of the double flux. 

Then, taking into account that the plasma of closed strings is dilute,
one can perform the summation over the grand canonical ensemble 
of these objects, that yields [instead of Eq.~(\ref{odinnad})] the following expression for the 
partition function:

\begin{equation}
\label{14new}
{\cal Z}=\int {\cal D}h_{\mu\nu}\exp
\left\{-\int d^4x\left[\frac{1}{24\eta^2}H_{\mu\nu\lambda}^2+
\frac{g_m^2}{4}h_{\mu\nu}^2-2\zeta\cos\left(\frac{\left|h_{\mu\nu}
\right|}{\Lambda^2}\right)\right]\right\}.
\end{equation} 
Here $\left|h_{\mu\nu}\right|\equiv\sqrt{h_{\mu\nu}^2}$, and 
$\Lambda\equiv\sqrt{\frac{L}{a^3}}$ is an UV momentum cutoff
with $L$ and $a$ denoting the characteristic distances between 
closed strings and their typical sizes, respectively.
Clearly, in the dilute-plasma approximation under study, 
$a\ll L$ and $\Lambda\gg a^{-1}$. Also in Eq.~(\ref{14new}),  
$\zeta\propto {\rm e}^{-S_0}$ stands for the fugacity
(Boltzmann factor) of a single string, which  
has the dimension $({\rm mass})^4$, with 
$S_0$ denoting the action of a single string.
The value of $S_0$ parametrically equals $\sigma a^2$,
where the area of the string world sheet is proportional to $a^2$, and 
$\sigma$ is the string tension;
$\sigma\simeq 2\pi\eta^2\ln\left(\frac{\lambda}{g_m^2}\right)$
in the London limit 
$\ln\left(\frac{\lambda}{g_m^2}\right)\gg 1$.

The square of the 
full mass of the field $h_{\mu\nu}$ following from Eq.~(\ref{14new}) 
reads $M^2=m^2+m_D^2\equiv Q^2\eta^2$. Here,    
$m_D^2=8\zeta\eta^2/\Lambda^4$ is the additional 
contribution, emerging due to the Debye screening of 
the dual vector boson in the plasma of closed strings, and 
$Q^2=2\left(g_m^2+\frac{4\zeta}{\Lambda^4}\right)$ 
is the (squared) full magnetic charge of the dual vector boson. 

To study the correlation functions of closed strings, it is convenient 
to represent the partition function~(\ref{14new}) directly as 
an integral over the densities of these objects. This can be done 
by means of some kind of a Legendre transformation, and the resulting action reads 

\begin{equation}
\label{15new}
S=2(\pi\eta)^2
\int d^4x\int d^4y \Sigma_{\mu\nu}(x)D_m^{(4)}(x-y)
\Sigma_{\mu\nu}(y)+V[\Sigma_{\mu\nu}],
\end{equation}
where the effective potential of closed strings, $V$, is

\begin{equation}
\label{potloops}
V[\Sigma_{\mu\nu}]=\int d^4x\left\{\Lambda^2|\Sigma_{\mu\nu}|\ln\left[
\frac{\Lambda^2}{2\zeta}|\Sigma_{\mu\nu}|+\sqrt{1+\left(
\frac{\Lambda^2}{2\zeta}|\Sigma_{\mu\nu}|\right)^2}\right]-2\zeta
\sqrt{1+\left(
\frac{\Lambda^2}{2\zeta}|\Sigma_{\mu\nu}|\right)^2}\right\}.
\end{equation}
It can be proved that the 
correlation functions of $\Sigma_{\mu\nu}$'s, 
evaluated by virtue of the representation~(\ref{15new}), are nothing, 
but the correlation functions of densities of closed strings in the plasma. 
These correlation functions 
can be calculated in the approximation when the plasma  
is sufficiently dilute, namely its density obeys the inequality
$\left|\Sigma_{\mu\nu}\right|\ll\frac{\zeta}{\Lambda^2}$, and 
the potential~(\ref{potloops}) becomes 
a simple quadratic functional of $\Sigma_{\mu\nu}$'s. 
In particular, the simplest nontrivial correlation function 
$\left<\left<\Sigma_{\mu\nu}(y)
\Sigma_{\lambda\rho}(y')\right>\right>_{x_\mu(\xi)}$
can be evaluated in this approximation. Inserting further 
the so-obtained expression for this correlation function into 
the average on the R.H.S. of Eq.~(\ref{otherhand}) (evaluated by means of the  
cumulant expansion in the bilocal approximation), one obtains for the functions 
${\cal D}$ and ${\cal D}_1$~\cite{10}:

\begin{equation}
\label{Dtot}
{\cal D}^{\rm full}\left(x^2\right)=\frac{m^2M}{4\pi^2}
\frac{K_1(M|x|)}{|x|},
\end{equation}

\begin{equation}
\label{D1tot}
{\cal D}_1^{\rm full}\left(x^2\right)=\frac{m_D^2}{\pi^2M^2|x|^4}+
\frac{m^2}{2\pi^2Mx^2}\left[\frac{K_1(M|x|)}{|x|}+\frac{M}{2}
\left(K_0(M|x|)+K_2(M|x|)\right)\right].
\end{equation}

We see that, as it should be, the functions~(\ref{Dtot}) and 
(\ref{D1tot}) go over into Eqs.~(\ref{dvadtri}) and (\ref{dvadchetyr}),
respectively, when $m_D\to 0$, {\it i.e.} when one neglects the effect of 
screening in the ensemble of closed strings. An obvious important 
consequence of the obtained Eqs.~(\ref{Dtot}) and (\ref{D1tot}) is that
the correlation length of the vacuum, $T_g$, becomes modified 
from $m^{-1}$ [according to Eqs.~(\ref{dvadtri}) and (\ref{dvadchetyr})]
to $M^{-1}$. (It is worth pointing out once again that this effect is 
due to the Debye screening of the dual vector boson 
in the ensemble of closed strings, that makes 
this particle heavier, namely its mass becomes increased from $m$ to $M$.)
Indeed, it is straightforward to see that, at $|x|\gtrsim M^{-1}$,

$${\cal D}^{\rm full}\longrightarrow\frac{(mM)^2}{4\sqrt{2}
\pi^{\frac32}}\frac{{\rm e}^{-M|x|}}{(M|x|)^{\frac32}},~~
{\cal D}_1^{\rm full}\longrightarrow
\frac{m_D^2}{\pi^2M^2|x|^4}+\frac{(mM)^2}{2\sqrt{2}\pi^{\frac32}}
\frac{{\rm e}^{-M|x|}}{(M|x|)^{\frac52}}.$$
A remarkable fact is   
that the leading term of the IR asymptotics 
of the function ${\cal D}_1^{\rm full}$ is a pure power-like one, rather 
than that of the function ${\cal D}_1$, given by Eq.~(\ref{dvadshest}).
This term produces a nonperturbative $(1/r)$-contribution to the $\bar qq$-potential, $\Delta V(r)=-\frac{(m_D/M)^2}{4\pi r}$,
which by its structure resembles the L\"uscher term. Typically, modelling  
the L\"uscher term within the SVM is rather problematic. Indeed, in the standard approach, in order to get
the L\"uscher term, one should consider 
string fluctuations, while SVM is well defined only on the minimal-area surface (see {\it e.g.} Ref.~\cite{2}).
Now, we have found another mechanism, which might generate a L\"uscher-type term 
via a novel nonperturbative perimeter interaction.

It is also worth noting that, despite the modification of the 
${\cal D}$-function,
the string tension of the open dual-string world sheet $\Sigma^{\rm e}$, 
$\sigma=4T_g^2\int d^2z{\cal D}\left(z^2\right)$ ({\it cf.} Ref.~\cite{23}), becomes 
modified only by means of the logarithm of the Landau-Ginzburg parameter. Indeed,
one obtains $\sigma=8\pi\eta^2\ln(\lambda/Q^2)\propto\eta^2$, and $\eta$ is not affected by 
the Debye screening. The screening rather modifies more significantly 
the coupling constant of the next-to-leading term in the derivative 
expansion of the nonlocal string effective action 
(the so-called rigidity term).
Indeed, by virtue of the results of Ref.~\cite{23}, one can see that,
for the same world sheet $\Sigma^{\rm e}$, this coupling constant 
without taking screening into account reads $-\frac{\pi}{2g_m^2}$,
whereas in the presence of screening it goes over to $-\frac{\pi}{2\left(g_m^2+\frac{4\zeta}{\Lambda^4}\right)}=-\frac{\pi}{Q^2}$.

Another origin of corrections to the $\bar q q$-potential (even without accounting for closed strings)
is due to the deviation from the London limit~\cite{koma}:

$$V(r)=-g^2\frac{{\rm e}^{-mr}}{4\pi r}\left[1-{\rm e}^{-\left(\sqrt{m^2+m_H^2}-m\right)r}+
{\rm e}^{-\left(m_H-m\right)r}\right], r>m_H^{-1}.$$
Clearly, this potential is neither Yukawa, nor Coulombic one, but it goes to the Yukawa potential in the 
London limit $m_H\to\infty$.

\section{String representation of the SU($N$)-inspired DAHM with the $\Theta$-term}

In this Section, we will present a string representation of the SU($N$)-inspired analogue of the model~(\ref{et5}),
extended, for completeness, by the $\Theta$-term. Owing to this term, quarks acquire a nonvanishing magnetic charge ({\it i.e.}, 
become dyons) and scatter off closed dual strings. As one of the consequences of our result, we will get the
critical values of $\Theta$, at which the long-range topological interaction of dual strings
with dyons disappears. These values, in particular, reproduce the respective SU(2)- and SU(3)-ones, found in 
Refs.~\cite{emil} and~\cite{theta}, respectively. The partition function of the
effective $[U(1)]^{N-1}$ gauge-invariant Abelian-projected theory
we are going to explore reads

$$
{\cal Z}_\alpha=\int\left(\prod\limits_{i}^{} \left|\Phi_i\right| {\cal D}\left|\Phi_i\right|
{\cal D}\theta_i\right) {\cal D}{\bf B}_\mu
\delta\left(\sum\limits_{i}^{}
\theta_i\right)\exp\Biggl\{-\int d^4x\Biggl[\frac14\left({\bf F}_{\mu\nu}+{\bf F}_{\mu\nu}^{(\alpha)}\right)^2+$$

\begin{equation}
\label{et6}
+\sum\limits_{i}^{}\left[\left|\left(\partial_\mu-
ig_m{\bf q}_i{\bf B}_\mu\right)\Phi_i\right|^2+
\lambda\left(|\Phi_i|^2-\eta^2\right)^2\right]-\frac{i\Theta g_m^2}{16\pi^2}
\left({\bf F}_{\mu\nu}+{\bf F}_{\mu\nu}^{(\alpha)}\right)
\left(\tilde{\bf F}_{\mu\nu}+\tilde{\bf F}_{\mu\nu}^{(\alpha)}\right)
\Biggr]\Biggr\}.
\end{equation}
Here, the index $i$ runs from 1 to the number of positive roots ${\bf q}_i$'s of the SU($N$)-group, that is $N(N-1)/2$.
Note that the origin of root vectors in Eq.~(\ref{et6}) is the fact that
monopole charges are distributed along them. Further,
$\Phi_i=\left|\Phi_i\right|{\rm e}^{i\theta_i}$ are the
dual Higgs fields, which describe the condensates of monopoles, and
${\bf F}_{\mu\nu}=\partial_\mu{\bf B}_\nu-\partial_\nu{\bf B}_\mu$ is the
field-strength tensor of the
$(N-1)$-component ``magnetic'' potential ${\bf B}_\mu$. The latter is dual
to the ``electric'' potential, whose components are diagonal gluons.
Since the SU($N$)-group is special, the phases $\theta_i$'s of the
dual Higgs fields are related to each other by the constraint
$\sum\limits_{i}^{}\theta_i=0$, which is imposed by introducing
the corresponding $\delta$-function into the R.H.S. of Eq.~(\ref{et6}) [{\it cf.} Ref.~\cite{suz} for the SU(3)-case].
Next, the index $\alpha$ runs from 1 to $N$ and denotes a certain quark colour.
Finally,
${\bf F}_{\mu\nu}^{(\alpha)}$ is the field-strength tensor
of a test quark of the colour $\alpha$, which moves along a certain
contour $C$. This tensor obeys the equation
$\partial_\mu\tilde {\bf F}_{\mu\nu}^{(\alpha)}=g{\bf m}_\alpha j_\nu$,
where $j_\mu(x)=\oint\limits_{C}^{}dx_\mu(\tau)\delta(x-x(\tau))$,
and ${\bf m}_\alpha$ is a weight vector of the fundamental representation of the group SU($N$).
One thus has ${\bf F}_{\mu\nu}^{(\alpha)}=-g{\bf m}_\alpha\tilde\Sigma_{\mu\nu}^{\rm e}$.
Note further that the $\Theta$-term can be rewritten as

\begin{equation}
\label{ch}
-\frac{i\Theta g_m^2}{16\pi^2}
\left({\bf F}_{\mu\nu}+{\bf F}_{\mu\nu}^{(\alpha)}\right)
\left(\tilde{\bf F}_{\mu\nu}+\tilde{\bf F}_{\mu\nu}^{(\alpha)}\right)=\frac{i\Theta g_m}{\pi}{\bf m}_\alpha
\int d^4x{\bf B}_\mu j_\mu.
\end{equation}
This means that, by means of the $\Theta$-term, quarks acquire a nonvanishing magnetic charge $\Theta g_m/\pi$, {\it i.e.}
become dyons, that enables them to interact with the magnetic gauge field ${\bf B}_\mu$~\cite{witten}.

Expanding for a while $|\Phi_i|$ around the Higgs {\it v.e.v.} $\eta$, one gets the mass of the dual vector boson,
$m=g_m\eta\sqrt{N}$. In what follows, we will again consider the London limit of the model~(\ref{et6}),
which admits a construction of the string representation. This is the limit when $m$ is much smaller
than the mass of any of the Higgs fields, $m_H=2\eta\sqrt{\lambda}$. Since we would like
our model to be consistent with QCD, we must have $g=\sqrt{\bar\lambda/N}$, where $\bar\lambda$
remains finite in the large-$N$ limit. Therefore, in the London limit,
the Higgs coupling $\lambda$ should grow with $N$ faster than ${\cal O}\left(N^2\right)$, namely it should obey the
inequality $\lambda\gg (2\pi N)^2/\bar\lambda$.

Integrating then $|\Phi_i|$'s out, we arrive at the following expression for the partition function~(\ref{et6})
in the London limit:

$$
{\cal Z}_\alpha=\int\left(\prod\limits_{i}^{}
{\cal D}\theta_i^{\rm sing.}{\cal D}\theta_i^{\rm reg.}\right) {\cal D}{\bf B}_\mu{\cal D}k
\delta\left(\sum\limits_{i}^{}
\theta_i^{\rm sing.}\right)\exp\Biggl\{-\int d^4x\Biggl[\frac14\left({\bf F}_{\mu\nu}+{\bf F}_{\mu\nu}^{(\alpha)}\right)^2+$$

\begin{equation}
\label{et7}
+\eta^2\sum\limits_{i}^{}\left(\partial_\mu\theta_i-
g_m{\bf q}_i{\bf B}_\mu\right)^2-ik\sum\limits_{i}^{}\theta_i^{\rm reg.}-\frac{i\Theta g_m^2}{16\pi^2}
\left({\bf F}_{\mu\nu}+{\bf F}_{\mu\nu}^{(\alpha)}\right)
\left(\tilde{\bf F}_{\mu\nu}+\tilde{\bf F}_{\mu\nu}^{(\alpha)}\right)
\Biggr]\Biggr\}.
\end{equation}
The multivalued fields $\theta_i^{\rm sing.}$'s here are related to the world sheets of closed dual strings $\Sigma_i$'s
by the same Eq.~(\ref{devyat}). The string representation of this partition function reads~\cite{suNrep}

$$
{\cal Z}_\alpha=
\exp\left\{-\frac{N-1}{4N}\left[g^2+\left(\frac{\Theta g_m}{\pi}\right)^2\right]
\int d^4x d^4y j_\mu(x)D_m(x-y)j_\mu(y)\right\}
\int\left(\prod\limits_{i}^{}{\cal D}x^{(i)}(\xi)\right)
\times$$

$$
\times\delta\left(\sum\limits_{i}^{}\Sigma_{\mu\nu}^i\right)
\exp\Biggl[
-2(\pi\eta)^2\int d^4x d^4y\hat\Sigma_{\mu\nu}^i(x)
D_m(x-y)\hat\Sigma_{\mu\nu}^i(y)-2i\Theta s_i^{(\alpha)}\hat L\left(\Sigma_i,C\right)+
$$

\begin{equation}
\label{main}
+2i\Theta\int d^4xd^4y\left(\frac{N-1}{N}\tilde\Sigma_{\mu\nu}(x)-s_i^{(\alpha)}\tilde\Sigma_{\mu\nu}^i(x)\right)j_\mu(y)
\partial_\nu^xD_m(x-y)\Biggr],
\end{equation}
where $\hat\Sigma_{\mu\nu}^i\equiv\Sigma_{\mu\nu}^i-Ns_i^{(\alpha)}\Sigma_{\mu\nu}$, 
and nonvanishing $s_i^{(\alpha)}$'s are equal $\pm N^{-1}$.
Note that,
for every color $\alpha$, it is straightforward to
integrate out one of
the world sheets $\Sigma_i$'s by resolving the constraint
imposed by the $\delta$-function.

The first exponent
on the R.H.S. of Eq.~(\ref{main}) represents the short-ranged interaction
of quarks via dual vector bosons. Noting that, for any $\alpha$,
${\bf m}_\alpha^2=(N-1)/(2N)$, one readily deduces from this term the total charge of the
quark, $\sqrt{g^2+(\Theta g_m/\pi)^2}$. The magnetic part of this charge coincides
with the one stemming from Eq.~(\ref{ch}).
Further, the first term in the
second exponent on the R.H.S. of Eq.~(\ref{main}) is again the
short-ranged (self-)interaction of
closed world sheets $\Sigma_i$'s and an open one $\Sigma$, responsible for confinement.
The last term on the R.H.S. of Eq.~(\ref{main}) describes
the short-range interactions
of dyons with both closed and open strings (obviously, the latter confine these very dyons themselves).
Instead, the term $-2i\Theta s_i^{(\alpha)}\hat L\left(\Sigma_i,C\right)$ in Eq.~(\ref{main})
describes the long-range
interaction of dyons with closed world sheets, that is the 4D-analogue of the Aharonov-Bohm effect~\cite{four}.
Since nonvanishing values of $s_i^{(\alpha)}$'s are equal $\pm N^{-1}$,
at $\Theta\ne N\pi\times{\,}{\rm integer}$,
dyons (due to their magnetic charge) do interact by means of this term with the closed dual strings.
On the contrary, these critical values of $\Theta$
correspond to such a relation
between the magnetic charge of a dyon and an electric flux inside the
string when the scattering of dyons off strings is absent.

\section{Summary}

In the present article, we have first briefly reviewed the properties of 
electric field-strength correlators in the DAHM, which
correspond to the gauge-invariant
correlators in the real QCD. First, we have reviewed the semi-classical
analysis of these correlators. Then, 
the leading correction to this result, produced by the interaction of the open-string 
world sheet with closed dual strings, has been evaluated. 
This effect is essentially quantum, as well as the plasma of closed strings
itself. In this way, it has been shown that the correlation length
of the vacuum becomes modified from the 
inverse mass of the dual vector boson, which it acquires by means of 
the Higgs mechanism, to its inverse full mass, which takes 
into account also the effect of Debye screening. What is more important is that, in one of the two
coefficient functions, which parametrize the bilocal correlator of 
electric field strengths within the SVM,  
a nonperturbative power-like IR part appears, which was absent on the semi-classical 
level. This novel term opens up a possibility of generating a
L\"uscher-type term within the SVM. We have further presented another type of modification of the 
$\bar q q$-potential, which appears beyond the London limit. The novel potential is a certain combination of 
Yukawa potentials with various effective masses, but it goes over to the standard Yukawa potential in the London limit.
Finally, we have discussed the string representation of the SU($N$)-counterpart of DAHM in the London limit, extended by the
$\Theta$-term. Owing to the latter, quarks have been shown to acquire a magnetic charge and scatter off closed dual strings,
provided $\Theta$ does not take its values from a certain discrete set.

In conclusion, the obtained results demonstrate similarities in the 
vacuum structures of DAHM and QCD by means of the SVM. 
They might also shed some light on the origin of the L\"uscher term
in QCD, as well as on the structure of the colour flux tubes.

\begin{acknowledgments}
One of the authors (D.A.) is grateful to the Alexander von Humboldt foundation for the financial support.
He would also like to thank the staff of the Institute of Physics of the Humboldt University of 
Berlin for cordial hospitality.
\end{acknowledgments}

\end{document}